\documentclass[a4paper, 10pt, aps, prl, reprint, nofootinbib, preprintnumbers]{revtex4-2}

\usepackage{physics}
\usepackage{amsmath, amssymb, amsfonts, dsfont}
\usepackage{hyperref}
\hypersetup{
colorlinks = true,
linkcolor = blue,
urlcolor = cyan,
citecolor = magenta
}
\usepackage{graphicx}
\usepackage{microtype}
\usepackage{orcidlink}

\begin{document}
\title{Tripartite nonlocal magic from reflected and negativity spectra}
\author{Sriram Akella\,\orcidlink{0000-0003-1557-8019}}
\email{sriram.akella@tifr.res.in}
\affiliation{Department of Theoretical Physics, Tata Institute of Fundamental Research, Homi Bhabha Road, Colaba, Mumbai 400005, India}
\date{\today}

\begin{abstract}
A tripartite pure state is local-unitary equivalent to a stabilizer state if and only if its marginal, reflected, and negativity spectra (up to signs) are flat. These conditions are quantified using anti-flatness, giving a diagnostic of tripartite nonlocal magic that is computable and experimentally measurable using randomized measurements. In non-abelian quantum doubles, it appears only when the triple points of the tripartition lie on plaquettes. In random tensor networks, it appears exactly when the entanglement wedge of two boundary regions connects.  
\end{abstract}
\preprint{TIFR/TH/26-35}

\maketitle

\textit{Introduction.} Stabilizer states~\cite{Gottesman:1997zz} are prepared by Clifford circuits and efficiently simulated on a classical computer~\cite{Gottesman:1998hu, Aaronson:2004xuh}. Nonstabilizerness, or \emph{magic}~\cite{Knill:2004ctr, Bravyi:2004isx}, becomes a resource \cite{Emerson:2013zse, Howard:2017maw} for quantum advantage and a measure of quantum complexity \cite{Liu:2020yso, Basu:2025mmm}. In a many-body system, much of this magic is local and not long range; it can be created or removed by unitaries acting within each subsystem~\cite{Jejjala:2026lgm}. This motivates \emph{nonlocal magic}~\cite{White:2020zoz, Cao:2023mzo, Cao:2024nrx, Qian:2025oit}. For a pure state on $\mathcal{H}_1\otimes\cdots\otimes\mathcal{H}_q$ and any magic measure $\mathcal{M}$, such as the stabilizer R\'enyi entropy \cite{Leone:2021rzd}, nonlocal magic is defined by minimizing over the local-unitary (LU) orbit:
\begin{equation}
    \mathrm{NLM}(\ket{\psi}) = \min_{U_1,\dots,U_q} \mathcal{M}\big(U_1\otimes\cdots\otimes U_q\ket{\psi}\big).
\end{equation}
It vanishes if and only if $\ket{\psi}$ is  LU-equivalent to a stabilizer state. 

For two parties, nonlocal magic is well understood. A bipartite pure state is LU-equivalent to a stabilizer state iff its entanglement spectrum is flat. Measures of spectral anti-flatness \cite{Tirrito:2023fnw} witness nonlocal magic, and recent works have computed the bipartite minimization in terms of the Schmidt spectrum \cite{Sierant:2026qji, Torre:2026pil, Franchini:2026dsg, Viscardi:2026oos, Sierant:2026lvg}. Beyond two parties no such simple characterization is available \cite{Kraus:2009vpy, Nest:2005edd}. Nonlocal magic can be related to operator entanglement~\cite{Andreadakis:2025mfw}, but not to simple invariants of the state. Inclusion--exclusion combinations of subsystem magic~\cite{Malvimat:2026oqf} are computable but not faithful. Having a computable and faithful diagnostic of nonlocal magic is also useful for holographic systems. Nonlocal magic is necessary for nontrivial area operators and for gravitational backreaction \cite{Cao:2023mzo, Cao:2024nrx, Cao:2026uoq}.

In this Letter we give a complete, spectral characterization of tripartite nonlocal magic. Our main result is that a pure state $\ket{\psi}_{ABC}$ is LU-equivalent to a stabilizer state if and only if
\begin{enumerate}
    \item[(a)] the marginals $\rho_A$, $\rho_B$, $\rho_C$ have flat spectra;
    \item[(b)] $\rho_{AB}$ has a flat \emph{reflected} spectrum, i.e.\ the canonical purification \cite{Dutta:2019gen} of $\rho_{AB}$ has a flat spectrum on $AA^\star$;
    \item[(c)] $\rho_{AB}$ has a flat \emph{negativity}~\cite{Vidal:2002zz} spectrum up to signs, i.e.\ the nonzero eigenvalues of the partial transpose \cite{Peres:1996dw, Horodecki:1996nc} $\rho_{AB}^{\Gamma}$ have the same absolute value.
\end{enumerate}
Necessity follows from the GHZ-extraction theorem for tripartite stabilizer states, which reduces them by local unitaries to GHZ states, Bell pairs, and product states~\cite{Bravyi:2005ztn, Looi:2011jrm, Wong:2025cwc}. We prove sufficiency below. 

Each condition can be quantified by a difference of R\'enyi entropies. For a normalized density matrix $\sigma$, define logarithmic anti-flatness \cite{Cao:2024nrx, Tirrito:2023fnw} as
\begin{equation}
    \mathcal{D}(\sigma) = \log \Tr\sigma^3 - 2 \log \Tr\sigma^2 \geq 0,
\end{equation}
which vanishes if and only if $\sigma$ has a flat spectrum. A tripartite diagnostic of nonlocal magic can then be built as a sum of $\mathcal{D}$'s associated to the above three conditions. Every term is the log of a multi-invariant~\cite{Gadde:2024taa, Akella:2026xza}, and can be evaluated by replica methods and measured with randomized-measurement protocols \cite{Huang:2020tih, Elben:2020hpu}.

We calculate this diagnostic in three examples. Non-abelian quantum doubles~\cite{Kitaev:1997wr, Bombin:2007qv} furnish states that satisfy (a) and (b) but not (c). On the lattice, Kitaev's ground state on a spherical lattice gives the same value at any system size when both triple points, where three regions meet, lie on plaquettes. A triple point on a vertex removes all nonlocal magic. In quantum doubles, tripartite nonlocal magic is a corner effect rather than an invariant of the phase. We then calculate the diagnostic in the vacuum of a two-dimensional conformal field theory and in random tensor networks~\cite{Hayden:2016cfa}, which model fixed-area states \cite{Dong:2018seb, Akers:2018fow, Bao:2018pvs} in holography. In random tensor networks the marginal spectra are flat and only the reflected and negativity terms survive. For two disjoint boundary arcs $A$ and $B$, both vanish when the entanglement wedge of $AB$ is disconnected, and switch on exactly when it is connected. Deep in the connected phase, the reflected term contributes a fixed amount for each endpoint of the entanglement wedge cross section, in parallel with the Markov gap \cite{Hayden:2021gno}. All technical details are collected in the supplemental material.

\textit{Necessity.} Condition (a) is a well-known fact about stabilizer states \cite{Fattal:2004frh}. It is the ingredient that goes into using anti-flatness as a diagnostic of bipartite nonlocal magic. Condition (b) is better known in a weaker form: stabilizer states have zero Markov gap, the difference between the reflected entropy and the mutual information~\cite{Hayden:2021gno, Zou:2020bly}. This follows from the GHZ-extraction theorem \cite{Bravyi:2005ztn, Akella:2025owv}. But it is also true that $\rho_{AB}$ coming from a tripartite stabilizer state $\ket{\psi}_{ABC}$ has a flat reflected spectrum, because both GHZ states and Bell pairs have a flat reflected spectrum. Condition (c) is less familiar, although partial-transpose moments of stabilizer states have been studied in~\cite{Carrasco:2022rcy}. While $\rho_{AB}$ coming from a GHZ state has a positive partial transpose with a flat spectrum, the partial transpose of a Bell pair has eigenvalues of equal magnitude and opposite signs. Tensor products of the two therefore satisfy (c).

\textit{Counterexamples.} Condition (a) by itself is not strong enough for sufficiency. For $d \geq 3$, the tripartite state \cite{Ramadas:2024fvd}
\begin{equation}\label{eq:ramadas}
   \frac{1}{d} \sum_{j, k = 0}^{d-1} e^{i \theta_{j, k}}\ket{j, k, j \oplus k}, \quad \theta_{j,k} = \begin{cases}
    \theta, & j = k = 0 \\
    0, & \text{otherwise}
    \end{cases}
\end{equation}
satisfies condition (a) for all values of $\theta$. However, for $\theta \neq 0$ it fails (b) and (c), and it cannot be LU-equivalent to a stabilizer state. Conditions (a) and (b) together are not strong enough either. For example, consider the ground state of the $S_3$ quantum double \cite{Kitaev:1997wr}: 
\begin{equation}\label{eq:qd}
    \ket{\psi_{S_3}} = \frac{1}{3!} \sum_{abc = e} \ket{a} \ket{b} \ket{c}, 
\end{equation}
where $a$, $b$, and $c$ are elements of the symmetric group $S_3$ satisfying the relation $abc = e$ (the identity permutation). It satisfies both conditions (a) and (b), but cannot be LU-equivalent to a stabilizer state as it fails condition (c). We return to non-abelian quantum doubles below.

\textit{Sufficiency.} Conditions (a) and (b) imply a vanishing Markov gap. A vanishing Markov gap forces $\ket{\psi}_{ABC}$ to be LU-equivalent to a direct sum of triangle states \cite{Zou:2020bly, Hayden:2003ytf}. The sectors of this direct sum may still differ in size, as they do for the quantum double state in Eq.~(\ref{eq:qd}). Condition (c) removes this freedom and forces all sectors to have equal weight. The resulting state is a tensor product of GHZ and Bell states. Therefore, only tensor products of stabilizer states satisfy conditions (a), (b), and (c).

\textit{Replica construction.} Condition (a) can be quantified using $\mathcal{D}(\rho_A) + \mathcal{D}(\rho_B)+\mathcal{D}(\rho_C)$. Condition (b) can be quantified using $\mathcal{D}^{R}(\rho_{AB}) = \mathcal{D}(\rho_{AA^{\star}})$, where $\rho_{AA^{\star}}$ is obtained by tracing out $BB^{\star}$ in the canonical purification of $\rho_{AB}$. The canonical purification of $\rho_{AB}$ is the pure state $\ket{\sqrt{\rho_{AB}}} \in \mathcal{H}_A \otimes \mathcal{H}_A^{\star} \otimes \mathcal{H}_B \otimes \mathcal{H}_B^{\star}$ in the doubled Hilbert space with the Hilbert-Schmidt norm. The square root can be made sense of by considering the $m \to 1/2$ limit of $\rho_{AB}^{m}$ as in \cite{Dutta:2019gen, Akers:2021pvd}. We instead use the fact that if $\rho_C$ has a flat spectrum, then $\rho_{AB}$ is automatically a projector and $\sqrt{\rho_{AB}} \propto \rho_{AB}$. We work with $\rho_{AA^{\star}} \propto \Tr_{B B^{\star}} \ketbra{\rho_{AB}}$ which has a replica interpretation. The proportionality constant is fixed by $\Tr \rho_{AA^{\star}} = 1$. When $\rho_C$ is not flat, $\mathcal{D}^R$ defined in this way differs from the anti-flatness of the true reflected spectrum. This does not affect the faithfulness of the diagnostic below, as it vanishes only when $\mathcal{D}(\rho_C) = 0$.

To quantify condition (c), consider 
\begin{equation}
    \theta_{AB} = \frac{\left(\rho_{AB}^{\Gamma}\right)^2}{\Tr \left(\rho_{AB}^{\Gamma}\right)^2}
\end{equation}
which is positive and has unit trace. $\rho_{AB}^{\Gamma}$ has a flat negativity spectrum iff $\theta_{AB}$ has a flat spectrum. Condition (c) can then be quantified using $\mathcal{D}^N(\rho_{AB}) = \mathcal{D}(\theta_{AB})$. Putting the three conditions together, we define the diagnostic 
\begin{equation}\label{eq:n3}
    \mathcal{N}_3 = \mathcal{D}(\rho_A) + \mathcal{D}(\rho_B) + \mathcal{D}(\rho_C) 
    + \mathcal{D}^{R}(\rho_{AB}) + \mathcal{D}^N({\rho_{AB}}).   
\end{equation}

Every term in the diagnostic is the logarithm of a multi-invariant: 
\begin{equation}\label{eq:multi-invariant}
    \mathcal{Z}(\sigma_1, \dots, \sigma_q) = \langle \psi|^{\otimes r} \Pi(\sigma_1) \otimes \dots \otimes \Pi(\sigma_q) \ket{\psi}^{\otimes r},
\end{equation}
defined by a choice of permutations $\sigma_1, \dots, \sigma_q \in \mathcal{S}_r$, where $r$ is called the replica number. The permutation $\sigma_i$ has a natural unitary action on $r$ replicas of the Hilbert space $\mathcal{H}_i$ represented by $\Pi(\sigma_i)$. For example, $\Tr(\rho_A^3)$ for a tripartite state $\ket{\psi}_{ABC}$ can be thought of as the multi-invariant with replica number $r = 3$ and permutations $\sigma_A = (123)$, and $\sigma_B = \sigma_C = (1)(2)(3)$ (the identity permutation). A more complicated example is $\Tr \rho_{AA^{\star}}^3$ which requires $r = 6$ replicas and permutations $\sigma_A = (12)(34)(56)$, $\sigma_B = (16)(23)(45)$, and $\sigma_C =$ identity. Similarly, $\Tr \theta_{AB}^3$ requires $r = 6$ replicas with $\sigma_A = \sigma_B^{-1} = (123456)$, and $\sigma_C =$ identity \cite{Calabrese:2012ew, Dong:2021clv}.

Each term in $\mathcal{N}_3$ is a combination of logarithms of multi-invariants, which are polynomials in the state bounded below by inverse powers of the local dimensions. This ensures the logarithm does not blow up and makes $\mathcal{N}_3$ continuous. States close to an LU-equivalent stabilizer state therefore have small $\mathcal{N}_3$. By compactness of the space of states, the converse also holds for fixed local dimensions. Therefore, $\mathcal{N}_3$ is robust for finite-dimensional systems. 

\textit{Non-abelian quantum doubles.} The example in Eq.~(\ref{eq:qd}) generalizes to any finite group $G$. A group Fourier transform on each party maps it to 
\begin{equation}\label{eq:psi_G}
    \ket{\psi_G} \simeq \bigoplus_\mu \frac{d_{\mu}}{\sqrt{|G|}}\ket{\Delta_\mu},
\end{equation}
where $\mu$ runs over the $N$ irreducible representations of $G$, with dimensions $d_\mu$, and $\ket{\Delta_\mu}$ is a triangle state of three maximally entangled pairs of dimension $d_\mu$. Every marginal has the flat spectrum $1/|G|$ and the reflected spectrum is flat, so (a) and (b) hold. The partial transpose has eigenvalues of magnitude $1/(|G|d_\mu)$ with multiplicity $d_\mu^4$, so
\begin{equation}\label{eq:ND}
    \mathcal{N}_3 = \mathcal{D}^N(\rho_{AB}) = \log\frac{\left(\sum_\mu d_\mu^2\right)\left(\sum_\mu d_\mu^{-2}\right)}{N^2}.
\end{equation}
This vanishes if and only if every $d_\mu = 1$, that is, if and only if $G$ is abelian. For example, $\mathcal{N}_3 = \log\frac{3}{2}$ for $S_3$ and $\log\frac{7}{3}$ for $A_4$.

\begin{figure}
    \centering
    \includegraphics[width=0.5\linewidth]{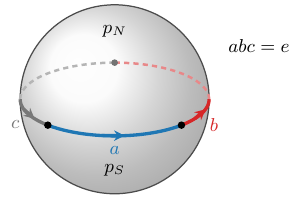}
    \caption{Quantum double ground state on the smallest spherical lattice. $a$, $b$, and $c$ are group elements satisfying the zero-flux condition $abc = e$ (the identity) on each plaquette. There are two triple points $p_N$ and $p_S$ where three regions come together. These lie on plaquettes of the lattice.}
    \label{fig:kitaev-sphere}
\end{figure}

\begin{figure}
    \centering
    \includegraphics[width=0.6\linewidth]{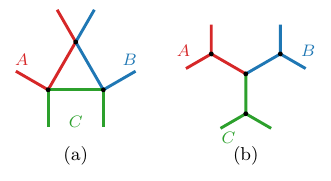}
    \caption{Triple points of a tripartition of a lattice into sets of edges $A$ (red), $B$ (blue), and $C$ (green). (a) A plaquette triple point, where no vertex touches all three parties. (b) A vertex triple point. For Kitaev's quantum double on the sphere, $\mathcal{N}_3$ is given by Eq.~(\ref{eq:ND}) if both triple points are of type (a) and vanishes otherwise.}
    \label{fig:corners}
\end{figure}

The state $\ket{\psi_G}$ is Kitaev's ground state on the smallest lattice on the sphere, with one edge per party (see Fig. \ref{fig:kitaev-sphere}). On a general spherical lattice, the ground state is 
\[
\sum_{h}\bigotimes_e\ket{h_{s(e)}h_{t(e)}^{-1}},\]
where $h_v \in G$ lives on the vertices and $s(e)$, $t(e)$ are the endpoints of edge $e$. Suppose the edges of each party fill a disk on the sphere, so there are exactly two triple points where the three regions come together ($p_N$ and $p_S$ in Fig. \ref{fig:kitaev-sphere}). We can use local unitaries to create maximally entangled pairs along each interface, which carry no magic. What remains are the two triple points. If both these points are at plaquettes as in Fig. \ref{fig:corners}(a), then the state is LU-equivalent to $\ket{\psi_G}$ in Eq. (\ref{eq:psi_G}), with $\mathcal{N}_3$ given by Eq. (\ref{eq:ND}) at any system size. However, if either triple point lies at a vertex as in Fig. \ref{fig:corners}(b), then the state is LU-equivalent to Bell pairs and a GHZ state over $G$, and $\mathcal{N}_3 = 0$. Tripartite nonlocal magic in quantum doubles is therefore not a property of the topological phase. A better probe of the phase is long-range magic, i.e., magic that cannot be removed by constant-depth local circuits \cite{Fliss:2020yrd, Korbany:2025noe, Wei:2025irp, Zhang:2026ckg, Korbany:2026fwo, Li:2026tsy}.

\textit{CFT$_2$ vacuum with two adjacent intervals.} In the vacuum of a two-dimensional CFT with adjacent intervals $A$ and $B$ of length $l_A$, and $l_B$, every term in $\mathcal{N}_3$ is a two- or three-point function of twist operators. Their dependence on the lengths is fixed by conformal invariance and the scaling dimensions of the twist operators \cite{Dixon:1986qv}. Putting together results from \cite{Calabrese:2009qy, Dutta:2019gen, Calabrese:2012ew}, we get
\begin{equation}\label{eq:n3-cft2}
\begin{split}
    \mathcal{D}(\rho_A) & = \frac{c}{18} \log \frac{l_A}{\epsilon}, \\ 
    \mathcal{D}^R(\rho_{AB}) &= \frac{c}{18}\log\frac{l_A\, l_B}{(l_A+l_B)\,\epsilon}, \\
    \mathcal{D}^N(\rho_{AB}) &= \mathcal{D}^R(\rho_{AB}) + \frac{c}{36}\log\frac{l_A+l_B}{\epsilon}.
\end{split}
\end{equation}
When $A$ and $B$ are not adjacent, we need to evaluate a $4$-point function of twist operators which is no longer universal. We instead appeal to random tensor networks.  

\textit{Random tensor networks.} Consider first a single random tensor $\ket{\psi}_{ABC}$ with the same local dimension $d_A = d_B = d_C = d \to \infty$.  Each multi-invariant is averaged over the Haar measure to give a sum over permutations \cite{Hayden:2016cfa}. To leading order in $1/d$, we find 
\begin{equation}\label{eq:n3-random}
\begin{split}
    & \mathcal{D}(\rho_A) \sim \frac{1}{d}, \, \, \,  \mathcal{D}^R(\rho_{AB}) \sim  \log d, \, \, \, \mathcal{D}^N(\rho_{AB}) = \log \frac{5}{4}. 
\end{split}
\end{equation}
The marginals are nearly flat and vanish in the $d \to \infty$ limit. Only the reflected term scales as $\log d$. The partial transpose obeys the semicircle law \cite{Shapourian:2020mkc, Dong:2021clv} whose moments are Catalan numbers giving $\mathcal{D}^N = \log(C_3 C_1/C_2^2)$. 

Following \cite{Akella:2026bci}, we next consider the random tensor network on a hyperbolic disk, truncated a distance $\epsilon$ from the boundary, with bond dimension $D \to \infty$. We work in the continuum limit where the tension between two domain walls carrying permutations $g$ and $h$ is $\kappa \, d(g, h)$, where $d$ is the Cayley distance. Partition the boundary into three adjacent arcs $A$, $B$, and $C$, of lengths $l_A$, $l_B$, and $l_C$. Each multi-invariant defines the boundary conditions for the partition function of the domain-wall spin model~\cite{Hayden:2016cfa, Penington:2022dhr}. The value of $\kappa = c/6$ is fixed by matching with the entanglement entropy of a CFT on a circle \cite{Ryu:2006bv}. 

\begin{figure}
    \centering
    \includegraphics[width=\linewidth]{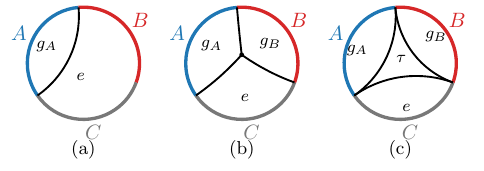}
    \caption{Dominant domain-wall saddles in the hyperbolic disk for (a) $\Tr\rho_A^n$, (b) $\Tr\rho_{AA^\star}^n$, and (c) $\Tr(\rho_{AB}^\Gamma)^{2n}$. Each region is labeled by its permutation: $g_A$ and $g_B$ are the boundary permutations $\sigma_A$ and $\sigma_B$ of the corresponding multi-invariant, and $e$ is the identity. In (c), $\tau$ lies on the Cayley-graph geodesic between each pair of boundary permutations.}
    \label{fig:rtn-saddles}
\end{figure}

Fig. \ref{fig:rtn-saddles} shows the bulk saddles that dominate in the large bond dimension limit for the R\'enyi entropies $\Tr \rho_A^n$, the reflected R\'enyi entropies $\Tr \rho_{AA^{\star}}^n$, and the partial transpose moments $\Tr \left(\rho_{AB}^{\Gamma}\right)^{2n}$. For the reflected moments, walls of tension $n$, $2n-2$, and $n$ meet at a bulk junction whose location is fixed by force balance. $\log[\Tr\rho_{AA^\star}^n/(\Tr\rho_{AB}^2)^n]$ is only sensitive to the contribution $-\kappa\, j(n)$ coming from the junction, where
\begin{equation}\label{eq:j(n)}
    j(n) = 2n\log n - (2n-1)\log(2n-1),
\end{equation}
while the cutoff and the arc lengths enter only through a term proportional to $n-1$, which drops out of $\mathcal{D}^{R}$. The corresponding anti-flatness terms in Eq. (\ref{eq:n3}) are
\begin{equation}\label{eq:rtn1}
\begin{split}
    & \mathcal{D}(\rho_A) = \mathcal{D}(\rho_B) = \mathcal{D}(\rho_C) = 0, \\
    & \mathcal{D}^R(\rho_{AB}) = \kappa\left[2j(2) - j(3)\right] = \frac{c}{6} \log \frac{2^8 5^5}{3^{12}}, \\
    & \mathcal{D}^N(\rho_{AB}) = \log\frac{5}{4}.
\end{split}
\end{equation}
As for the single random tensor, the reflected term scales with $c$, while the marginals are flat and the negativity term is $\mathcal{O}(1)$. 

\begin{figure}
    \centering
    \includegraphics[width=\linewidth]{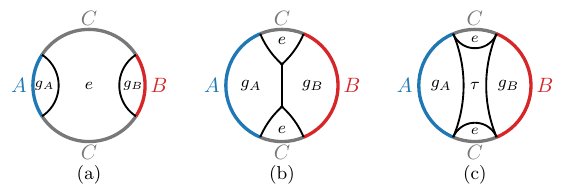}
    \caption{Dominant domain-wall saddles when $A$ and $B$ are disjoint arcs. (a) For $x < 1/2$, every moment is dominated by the disconnected saddle. (b) For $x > x_3$, the reflected moments are dominated by a $t$-channel network. (c) For $x > 1/2$, the negativity moments are dominated by $\tau$ filling the region between the four geodesics. Regions are labeled as in Fig.~\ref{fig:rtn-saddles}.}
    \label{fig:rtn-two-arcs}
\end{figure}

\begin{figure}
    \centering
    \includegraphics{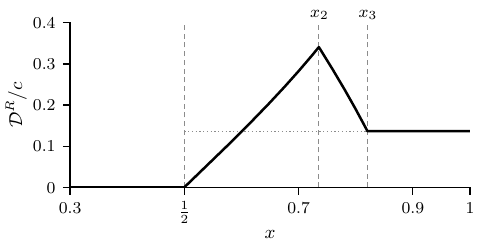}
    \caption{Anti-flatness of the reflected spectrum, $\mathcal{D}^R/c$, as a function of the cross-ratio $x$ for disjoint arcs, at leading order. The values $x_2 \approx 0.735$ and $x_3 \approx 0.821$ are determined numerically. The dotted line marks the plateau $\frac{1}{3}\log(2^8 5^5/3^{12})$.}
    \label{fig:rtn-DR}
\end{figure}

Next, let $A$ and $B$ be disjoint arcs, so that $C = C_1 \cup C_2$ consists of the two arcs between them. Label the endpoints $z_1, \dots, z_4$ in cyclic order, with $A = [z_1, z_2]$ and $B = [z_3, z_4]$, and let $z_{ij}$ be the chord length between $z_i$ and $z_j$. The cross-ratio $x = z_{12}z_{34}/(z_{13}z_{24})$ lies between $0$ and $1$, and the entanglement wedge of $AB$ is connected for $x > 1/2$ and disconnected otherwise~\cite{Headrick:2010zt}. Fig.~\ref{fig:rtn-two-arcs} shows the dominant saddles. For $x > 1/2$, the partial transpose moments are dominated by a $\tau$ filling the region bounded by the four geodesics, and we find 
\begin{equation}\label{eq:DN-rtn}
    \mathcal{D}^{N}(\rho_{AB}) = \log \frac{5}{4} \, \Theta(x - 1/2).
\end{equation}
The negativity spectrum is not flat precisely when the entanglement wedge is connected. The reflected moments are instead dominated at large $x$ by a \emph{$t$-channel} network, in which two junctions of the type found above are joined by a wall of tension $2n-2$ across the entanglement wedge cross section. This delays the transition of the $n$-th reflected moment to a cross-ratio $x_n > 1/2$, with $x_2 \approx 0.735$ and $x_3 \approx 0.821$ determined numerically. When $1/2 < x < x_2$, the reflected moment is in the disconnected phase while the normalization $\Tr \rho_{AB}^2$ is in the connected phase. At leading order the normalized moments are therefore $\Tr\rho_{AA^\star}^n = e^{-n I(A:B)}$, and $\mathcal{D}^R = -3I + 4I = I(A{:}B)$. Altogether,
\begin{equation}\label{eq:DR-two-arcs}
    \mathcal{D}^{R}(\rho_{AB}) = \begin{cases}
        0, & x < \frac{1}{2} \\
        I(A:B) = \frac{c}{3} \log \frac{x}{1-x}, & \frac{1}{2} < x < x_2, \\ 
        \frac{c}{3} \log \frac{2^8 5^5}{3^{12}}, & x > x_3,
    \end{cases}
\end{equation}
decreasing monotonically between $x_2$ and $x_3$ as shown in Fig.~\ref{fig:rtn-DR}. Beyond $x_3$, $\mathcal{D}^R$ is twice the three-arc value in Eq. (\ref{eq:rtn1}), one junction for each endpoint of the entanglement wedge cross section, just as the Markov gap of holographic states is at least $\frac{c}{3}\log 2$ per endpoint \cite{Hayden:2021gno}. The replica-dependent transition is familiar from the reflected entropy in random tensor networks \cite{Akers:2021pvd}. 

For disjoint arcs, $\mathcal{N}_3$ is nonzero only when the entanglement wedge of $AB$ is connected. This is also precisely when there is a Python's lunch \cite{Brown:2019rox, Engelhardt:2021mue} for $AB$. Although there is a Python's lunch for the region $C = C_1 \cup C_2$ when $x < 1/2$, local unitaries on $C$ can act jointly on $C_1$ and $C_2$ to ``digest" the Python's lunch, giving $\mathcal{N}_3 = 0$. For $x > 1/2$, the Python's lunch of $AB$ is not digested as we are not allowed to jointly act on $A$ and $B$, and we find a nonzero $\mathcal{N}_3$. This suggests a connection between nonlocal magic and the Python's lunch, in line with the recent observation that the Wigner negativity of a holographic state is set by the size of its Python's lunch~\cite{Bhattacharya:2026bda}. 

\textit{Discussion.} We showed that a tripartite pure state has no nonlocal magic if and only if its marginal, reflected, and negativity spectra (up to signs) are flat. The resulting diagnostic $\mathcal{N}_3$ needs at most $r = 6$ replicas and can be estimated with randomized measurements \cite{Huang:2020tih, Elben:2020hpu}. In quantum doubles, it is nonzero only when both triple points lie on plaquettes. In random tensor networks with disjoint arcs $A$ and $B$, it switches on exactly when the entanglement wedge of $AB$ connects. 

We close with some questions for future work. Can we go beyond three parties? In an evaporating black hole, the connectivity of the entanglement wedge changes at the Page time~\cite{Page:1993wv, Penington:2019npb, Almheiri:2019psf}, and the Wigner negativity of the radiation becomes large past the Page transition~\cite{Basu:2025uxw}. Does $\mathcal{N}_3$ among the black hole, the early radiation, and late radiation turn on at the Page time? In the many-body setting, bipartite nonlocal magic spreads very differently in chaotic and integrable dynamics~\cite{Aditya:2026qna}. How does $\mathcal{N}_3$ evolve in a random or Clifford+T circuit? Is there a measurement-induced phase transition, as there is for magic~\cite{Niroula:2023meg, Bejan:2023zqm}? What is $\mathcal{N}_3$ in a large-$c$ CFT \cite{Chandra:2022bqq} for two disjoint arcs? 

\textit{Acknowledgments.} I thank my PhD advisor, Abhijit Gadde, for support and guidance. I also thank Onkar Parrikar and Pratik Rath for useful discussions. This work is supported by the Department of Atomic Energy, Government of India, under Project Identification No. 4012, and the Infosys Endowment for the study of the Quantum Structure of Spacetime. 

\textit{AI Usage Statement.} I used Claude Opus 5.5 (Anthropic) in preparing this draft. \textbf{Figures.} The LLM generated the TikZ code for all figures. \textbf{Language and Error-Checking.} The LLM polished sentences, improved logical flow, corrected typos, and checked for errors/gaps in the draft. \textbf{Research Content.} The LLM constructed the two counterexamples in Eqs. \eqref{eq:ramadas} and \eqref{eq:qd} and pointed me to Ref. \cite{Ramadas:2024fvd}. The quantum double example, in particular, highlighted why marginal flatness and reflected flatness were not sufficient. I formulated condition (c), proved sufficiency, and worked out the three examples. No AI is an author and I take full responsibility for the content of the paper. 

\bibliography{refs}

\clearpage

\onecolumngrid
\appendix

\section*{Supplemental Material}

\setcounter{equation}{0}
\renewcommand{\theequation}{S\arabic{equation}}

This Supplemental Material collects the derivations behind the results quoted in the main text. Equation numbers prefixed with an S refer to this document; all other equation numbers refer to the main text. We first show that flat marginal and reflected spectra imply a zero Markov gap, a fact we used in the main text without proof. Then we work out the sketched sufficiency proof from the main text in more detail. We then provide more details about the quantum double example, the CFT\textsubscript{2} vacuum with two adjacent intervals, the single random tensor, and the random tensor network on the hyperbolic disk. 

\subsection{Flat spectra imply a vanishing Markov gap}

We show that if the marginals $\rho_A$, $\rho_B$, $\rho_C$ and the reflected spectrum of $\rho_{AB}$ are flat, then the Markov gap 
\begin{equation} 
h = S_R(A{:}B) - I(A{:}B)
\end{equation}
vanishes. Let $r_A$, $r_B$, and $r_C$ be the ranks of the marginals $\rho_{A}$, $\rho_{B}$, and $\rho_C$. From Schmidt decomposition, it follows that 
\begin{equation}
    \rho_{AB} = \frac{1}{r_C} P, 
\end{equation}
where $P$ is a projector onto the support of $\rho_{AB}$ satisfying $P^2 = P$, and $\Tr P = r_C$. The canonical purification of $\rho_{AB}$ is therefore a vectorization of the projector $P$, and can be Schmidt decomposed across the $AA^{\star}|BB^{\star}$ partition. This gives us 
\begin{equation}
    \ket{\sqrt{\rho_{AB}}} = \frac{1}{\sqrt{r_C}} \sum_{i = 1}^{R} s_i \, \ket{X_i}_{AA^{\star}} \otimes \ket{Y_i}_{BB^{\star}}.  
\end{equation}
The eigenvalues of $\rho_{AA^{\star}}$ are therefore $s_i^2/r_C$, and flatness implies that all $s_i$'s are equal to a common value $s$. The fact that $P$ is a projector means that $\Tr P^2 = \Tr P$, and using the above Schmidt decomposition of $P$, we conclude that 
\begin{equation}\label{eq:S4}
    R s^2 = r_C. 
\end{equation}

Next, let $\Pi_A$ and $\Pi_B$ be the projectors onto the support of $\rho_A$ and $\rho_B$, i.e., 
\begin{equation}
    \rho_A = \frac{1}{r_A} \Pi_A, \quad \rho_B = \frac{1}{r_B} \Pi_B. 
\end{equation}
The support of $P$ is contained in the support of $\Pi_A \otimes \Pi_B$, i.e., $P \leq \Pi_A \otimes \Pi_B$. To see this, note that $\Tr[((\mathds{1}-\Pi_A)\otimes\mathds{1})\,\rho_{AB}] = \Tr[(\mathds{1}-\Pi_A)\rho_A] = 0$. Since $\rho_{AB} \geq 0$, this forces $((\mathds{1}-\Pi_A)\otimes \mathds{1})\,\rho_{AB} = 0$, and similarly for $B$. Therefore
\begin{equation}
    \Tr\left[P\,(\Pi_A \otimes \Pi_B)\right] = \Tr P = r_C.
\end{equation}
The vectorization of $(\Pi_A \otimes \Pi_B)/\sqrt{r_A r_B}$ is a normalized product state across the $AA^{\star}|BB^{\star}$ partition, and the overlap of $\ket{P}$ with any normalized product state is at most its largest Schmidt coefficient. Hence
\begin{equation}
    \Tr\left[P \left(\frac{\Pi_A}{\sqrt{r_A}} \otimes \frac{\Pi_B}{\sqrt{r_B}}\right)\right] \leq s \implies s \geq \frac{r_C}{\sqrt{r_A r_B}}.
\end{equation}
From Eq. (\ref{eq:S4}), we get
\begin{equation}
    R \leq \frac{r_A r_B}{r_C} \implies S_R(A:B) \leq I(A: B). 
\end{equation}
But we know that $S_R(A:B) \geq I(A:B)$ \cite{Dutta:2019gen, Hayden:2021gno}, and therefore we conclude that the Markov gap is zero.

\subsection{Proof of sufficiency of conditions (a), (b), and (c)}

We now turn to the sufficiency proof sketched in the main text and fill in the details. As we just showed, conditions (a) and (b) imply a vanishing Markov gap. This forces the state to be a direct sum over triangle states \cite{Zou:2020bly, Hayden:2003ytf}. We now show that conditions (a) and (c) remove the freedom of freely choosing weights in the sum, leaving behind a tensor product of stabilizer states. 

Recall that a triangle state is a tensor product of the form: 
\begin{equation}
    \ket{\Delta} = \ket{\phi}_{A_1 B_1} \otimes \ket{\xi}_{B_2 C_1} \otimes \ket{\chi}_{A_2 C_2}. 
\end{equation}
The theorem of \cite{Zou:2020bly, Hayden:2003ytf} states that a state has zero Markov gap iff it is a direct sum of such triangle states: 
\begin{equation}\label{eq:S10}
    \ket{\Psi}_{ABC} = \bigoplus_{\mu} \sqrt{p_{\mu}} \, \ket{\Delta_{\mu}},
\end{equation}
with $\ket{\Delta_{\mu}}$ a triangle state in the sector $\mu$. The probabilities $p_{\mu}$ are arbitrary. We start with such a state and impose the flatness conditions (a) and (c). 

The marginal density matrix $\rho_A$ is 
\begin{equation}
    \rho_A = \bigoplus_{\mu} p_{\mu}\,  \rho_{A_1}^{(\mu)} \otimes \rho_{A_2}^{(\mu)}, 
\end{equation}
where $\rho_{A_1}^{(\mu)}$ and $\rho_{A_2}^{(\mu)}$ are the reduced density matrices coming from $\ket{\phi}$ and $\ket{\chi}$ in each sector. A flat spectrum for $\rho_A$ implies that every $\rho_{A_1}^{(\mu)}$ and $\rho_{A_2}^{(\mu)}$ is flat. Let $m_{\mu}$, $n_{\mu}$, and $k_{\mu}$ be the Schmidt ranks of $\ket{\phi}$, $\ket{\xi}$, and $\ket{\chi}$ in the sector $\mu$. The eigenvalues of the above $\rho_A$ are $p_{\mu}/m_{\mu}k_{\mu}$. Similarly, flatness of $\rho_B$ and $\rho_C$ imply 
\begin{equation}
    \frac{p_{\mu}}{m_{\mu}k_{\mu}} = \frac{1}{r_A}, \quad \frac{p_{\mu}}{m_{\mu}n_{\mu}} = \frac{1}{r_B}, \quad \frac{p_{\mu}}{n_{\mu} k_{\mu}} = \frac{1}{r_C}. 
\end{equation}
Solving for $m_{\mu}$, $n_{\mu}$, and $k_{\mu}$ in terms of $p_{\mu}$ gives 
\begin{equation}\label{eq:S13}
    m_{\mu}^2 = p_{\mu}\frac{r_A r_B}{r_C}, \quad n_{\mu}^2 = p_{\mu} \frac{r_B r_C}{r_A}, \quad k_{\mu}^2 = p_{\mu} \frac{r_A r_C}{r_B}. 
\end{equation}
Therefore, any set of $\{p_{\mu}\}$ is allowed, as long as the triangle states carry ranks proportional to $\sqrt{p_{\mu}}$. As we will see later, the quantum double example realizes this precisely because $r_A = r_B = r_C = |G|$, $p_{\mu} = d_{\mu}^2/|G|$, and $m_{\mu} = n_{\mu} = k_{\mu} = d_{\mu}$, where $d_{\mu}$ are the dimensions of the irreducible representations of $G$. 

There is no need to impose condition (b) as it gives nothing new. The role of this constraint in the proof is to establish the sum of triangle states structure. 
Next, we impose condition (c). The reduced density matrix on $AB$ is 
\begin{equation}
    \rho_{AB} = \bigoplus_{\mu}\, p_{\mu} \, \ketbra{\phi^{(\mu)}}_{A_1 B_1} \otimes \rho_{A_2}^{(\mu)} \otimes \rho_{B_2}^{(\mu)}, 
\end{equation}
where $\rho_{A_2}^{(\mu)}$ and $\rho_{B_2}^{(\mu)}$ are flat with Schmidt ranks $k_{\mu}$ and $n_{\mu}$. A partial transpose does nothing to them but converts the Bell state on $A_1 B_1$ into a \emph{flip} operation. This is explained schematically by 
\begin{equation}
    \sum_{i,j = 1}^{m_{\mu}} \ket{i, i}\bra{j, j} \rightarrow \sum_{i, j = 1}^{m_{\mu}} \ket{i ,j} \bra{j, i}. 
\end{equation}
The flip operator $F$ has eigenvalues $\pm 1$, so the partially transposed Bell state $F/m_{\mu}$ has eigenvalue $1/m_{\mu}$ with multiplicity $m_{\mu}(m_{\mu}+1)/2$ and eigenvalue $-1/m_{\mu}$ with multiplicity $m_{\mu}(m_{\mu}-1)/2$. These are the dimensions of the symmetric and antisymmetric subspaces. Therefore, the eigenvalues of $\rho_{AB}^{\Gamma}$ are 
\[
\pm \frac{p_{\mu}}{m_{\mu} n_{\mu} k_{\mu}} = \pm \frac{1}{\sqrt{p_{\mu} r_A r_B r_C}} 
\]
from Eq. (\ref{eq:S13}). If $\rho_{AB}^{\Gamma}$ has a flat spectrum up to signs, we must have all $p_{\mu} = 1/N$, where $N$ is the number of sectors in the direct sum. From Eq. (\ref{eq:S13}), we also find that all the Schmidt ranks are independent of the sector, i.e., $m_{\mu} = m$, $n_{\mu} = n$, and $k_{\mu} = k$. 

Putting all these results together in Eq. (\ref{eq:S10}), we get 
\begin{equation}
    \ket{\Psi}_{ABC} = \bigoplus_{\mu = 1}^{N} \frac{1}{\sqrt{N}} \ket{\Phi_m}_{AB} \otimes \ket{\Phi_n}_{BC} \otimes \ket{\Phi_k}_{AC},
\end{equation}
where 
\begin{equation}
    \ket{\Phi_d} = \frac{1}{\sqrt{d}}\sum_{i = 1}^{d} \ket{ii}
\end{equation}
is the maximally entangled Bell state of Schmidt rank $d$. Since the sector weights and the Schmidt ranks $m$, $n$, and $k$ are independent of $\mu$, we can identify the sectors with an $N$-dimensional register within each party's support and write 
\begin{equation}
    \ket{\Psi}_{ABC} = \ket{\rm GHZ_N} \otimes  \ket{\Phi_m}_{AB} \otimes \ket{\Phi_n}_{BC} \otimes \ket{\Phi_k}_{AC}
\end{equation}
up to local unitaries. This completes the proof that any state satisfying conditions (a), (b), and (c) must be a tensor product of GHZ and Bell states. 

\subsection{Quantum double on the smallest spherical lattice} 
Let's now turn to the quantum double example introduced in the main text. To begin with, we consider the simplest case of the smallest lattice on the sphere shown in Fig. \ref{fig:kitaev-sphere}, and the ground state is 
\begin{equation}\label{eq:S19}
    \ket{\psi_G} = \frac{1}{|G|} \sum_{abc = e} \ket{a} \ket{b} \ket{c}. 
\end{equation}
Let's now discuss group Fourier transforms. Let $D^{\mu}$ be the unitary irreps of $G$ with dimension $d_{\mu}$. Define the state
\begin{equation}
    \ket{\mu, i, j} = \sqrt{\frac{d_{\mu}}{|G|}} \sum_{g \in G} D^{\mu}_{ij}(g) \ket{g}, 
\end{equation}
which acts as the Fourier transform of the state $\ket{g}$. From the Schur orthogonality relations, $\sum_{g} \overline{D^{\mu}_{ij}(g)}\, D^{\nu}_{kl}(g) = \frac{|G|}{d_{\mu}}\delta_{\mu\nu}\delta_{ik}\delta_{jl}$, it follows that
\begin{equation}
    \bra{\mu, i, j} \nu, k, l\rangle = \delta_{\mu\nu} \delta_{ik} \delta_{jl},
\end{equation}
and since $\sum_{\mu} d_{\mu}^2 = |G|$, these states form an orthonormal basis of $\mathbb{C}[G]$.
The group delta function in this basis can be written as 
\begin{equation}
    \delta_{g, e} = \frac{1}{|G|} \sum_{\mu} d_{\mu} \chi_{\mu}(g), 
\end{equation}
where $\chi_{\mu} = \Tr D^{\mu}$ is the character. Therefore, we have 
\begin{equation}
\begin{split}
    \ket{\psi_G} & = \frac{1}{|G|^2} \sum_{\mu} d_{\mu} \sum_{ijk} \left(\sum_{a} D^{\mu}_{ij}(a) \ket{a}\right) \left(\sum_b D_{jk}^{\mu}(b) \ket{b} \right) \left(\sum_c D^{\mu}_{ki}(c) \ket{c}\right) \\ 
    & = \frac{1}{\sqrt{|G|}} \sum_{\mu} \frac{1}{\sqrt{d_{\mu}}} \sum_{ijk} \ket{\mu, i, j}_A \ket{\mu, j, k}_B \ket{\mu, k, i}_C. 
\end{split}
\end{equation}
Let's define the normalized triangle state
\begin{equation}
    \ket{\Delta_{\mu}} = \frac{1}{d_{\mu}^{3/2}} \sum_{ijk} \ket{\mu, i, j}_A \ket{\mu, j, k}_B \ket{\mu, k, i}_C,
\end{equation}
and the ground state becomes 
\begin{equation}
    \ket{\psi_G} = \bigoplus_{\mu} \frac{d_{\mu}}{\sqrt{|G|}} \ket{\Delta_{\mu}}, 
\end{equation}
where the normalization is fixed by $\sum d_{\mu}^2 = |G|$. This proves Eq. (\ref{eq:psi_G}) from the main text. 

The state is a direct sum of triangle states with $p_{\mu} = d_{\mu}^2/|G|$ and all Schmidt ranks $m_{\mu} = n_{\mu} = k_{\mu} = d_{\mu}$. Therefore the eigenvalues of $\rho_A$, $\rho_B$, and $\rho_C$ are all $p_{\mu}/d_{\mu}^2 = 1/|G|$, and $r_A = r_B = r_C = |G|$. From our discussion of the partial transpose, the eigenvalues are $\pm p_{\mu}/d_{\mu}^3 = \pm 1/|G|d_{\mu}$ with multiplicity $d_{\mu}^3(d_{\mu} \pm 1)/2$. Therefore, the moments are
\begin{equation}
    N_2 = \sum_{\mu} \frac{d_{\mu}^4}{|G|^2 d_{\mu}^2} = \frac{1}{|G|}, \quad N_4 = \sum_{\mu} \frac{d_{\mu}^4}{|G|^4d_{\mu}^4} = \frac{N}{|G|^4}, \quad N_6 = \sum_{\mu} \frac{d_{\mu}^4}{|G|^6 d_{\mu}^6} = \frac{1}{|G|^6} \sum_{\mu} d_{\mu}^{-2},
\end{equation}
where $N$ is the total number of irreps, and $N_{k} = \Tr \left(\rho_{AB}^{\Gamma}\right)^{k}$ are the moments of the partial transpose.  The negativity term in $\mathcal{N}_3$ is 
\begin{equation}
    \mathcal{D}^N(\rho_{AB}) = \log N_6 - 2\log N_4 + \log N_2 = \log \frac{|G| \sum_{\mu} d_{\mu}^{-2}}{N^2} = \log \frac{\sum_{\mu} d_{\mu}^2 \sum_{\mu} d_{\mu}^{-2}}{N^2}, 
\end{equation}
which proves Eq. (\ref{eq:ND}). From the Cauchy--Schwarz inequality, this is zero iff all $d_{\mu}$ are equal, and as the trivial irrep always has $d_{\mu} = 1$, we conclude that $G$ must be abelian. 

\subsection{Quantum double on a general spherical lattice}

We now turn to a more general spherical lattice. Assign a group element $h_v \in G$ to each vertex $v$. Up to normalization, the ground state is
\begin{equation}
    \ket{\psi_{G}} \propto \sum_{\{h_v\}} \bigotimes_{e} \ket{h_{s(e)}h^{-1}_{t(e)}},
\end{equation}
where $s(e)$ and $t(e)$ are the \emph{source} and \emph{target} vertices of the edge $e$. Multiplying every $h_v$ on the right by the same group element does not change any edge, so only the vertex elements relative to one another are physical. We now explain the argument below Eq. (\ref{eq:ND}) in the main text. Let the edges of each party fill a disk on the sphere, so that there are exactly two triple points where the three regions meet, as shown in Figs. \ref{fig:kitaev-sphere} and \ref{fig:corners}.

Since the edges of a party are connected, the party can read off from its own edges the group elements on all the vertices it touches, relative to one another. A local unitary therefore lets each party store these vertex elements instead of its edges. Vertices that touch only one party are known to that party alone and can be disentangled. Vertices on the interface between two parties, say $A$ and $B$, are known to both. Their elements relative to one another are summed over freely and give maximally entangled Bell states of the form
\[
\sum_{g \in G} \ket{g}_A \otimes \ket{g}_B, 
\]
which are stabilizer states. What remains is how the three interfaces sit relative to each other, and this is decided by the triple points.

When both triple points sit at plaquettes, as in Fig. \ref{fig:corners}(a), no vertex touches all three parties. Party $A$ touches the $AB$ and $AC$ interfaces and knows a single group element $a$ that relates the two. Similarly, $B$ knows the element $b$ relating the $AB$ and $BC$ interfaces, and $C$ knows the element $c$ relating the $BC$ and $AC$ interfaces. Going once around all three interfaces brings us back to where we started, so $abc = e$, just like the zero-flux condition on the smallest lattice. Apart from this constraint, $a$, $b$, and $c$ are summed over freely, and the residual state is precisely the one in Eq. (\ref{eq:S19}). The two plaquettes together leave behind a single copy of this state, so $\mathcal{N}_3$ is given by Eq. (\ref{eq:ND}) at any system size.

If instead a triple point sits at a vertex, as in Fig. \ref{fig:corners}(b), then all three parties know the group element on this vertex and can measure everything relative to it. No relative element is left over, and the state is LU-equivalent to Bell pairs alone. If the second triple point is also a vertex, its element relative to the first is known to all three parties, which gives an additional GHZ state $\sum_g \ket{g}\ket{g}\ket{g}$. Either way, $\ket{\psi_G}$ is LU-equivalent to a tensor product of stabilizer states and has no tripartite nonlocal magic.

\subsection{\texorpdfstring{CFT\textsubscript{2}}{CFT2} vacuum with two adjacent intervals}
Take the vacuum of a CFT on the infinite line with $A = [-l_A,0]$, $B=[0,l_B]$, and $C$ the complement, with UV cutoff $\epsilon$. A multi-invariant with permutations $(\sigma_A,\sigma_B,\sigma_C=e)$ on $r$ replicas is a correlation function of twist operators in the orbifold theory $\mathrm{CFT}^{\otimes r}/S_r$~\cite{Calabrese:2009qy, Dixon:1986qv}. At each point where the permutation jumps from $\sigma_X$ on the left to $\sigma_Y$ on the right we insert a twist $\mathcal{T}_{\sigma_X^{-1}\sigma_Y}$. The twist operator $\mathcal{T}_\sigma$ is a primary with scaling dimension
\begin{equation}
    h = \bar{h} = \frac{c}{24}\sum_{i} \left(l_i - \frac{1}{l_i}\right),
\end{equation}
where $l_i$ are the cycle lengths of $\sigma$. For adjacent $A$ and $B$, there are only three insertion points at $x_1 = -l_A$, $x_2 = 0$, and $x_3 = l_B$, and the three-point function of twist operators is fixed up to a constant:
\begin{equation}\label{eq:S-3pt}
    \Big\langle \mathcal{T}_{\sigma_1}(x_1) \mathcal{T}_{\sigma_2}(x_2) \mathcal{T}_{\sigma_3}(x_3) \Big\rangle = C_{123}\left(\frac{x_{12}}{\epsilon}\right)^{-2(h_1+h_2-h_3)} \left(\frac{x_{23}}{\epsilon} \right)^{-2(h_2 + h_3 - h_1)}\left(\frac{x_{13}}{\epsilon} \right)^{-2(h_1 + h_3 - h_2)},
\end{equation}
where $x_{12} = l_A$, $x_{23} = l_B$, and $x_{13} = l_A + l_B$. The constants $C_{123}$, as well as the normalizations of the twist operators, are non-universal. They contribute $l$-independent constants to $\mathcal{N}_3$, which we drop.

For a single interval, we have 
\[
\Tr \rho_A^n = \langle \mathcal{T}_{\tau}(x_1) \mathcal{T}_{\tau^{-1}}(x_2) \rangle = c_n \left( \frac{x_{12}}{\epsilon}\right)^{-4 h}, 
\]
where $\tau$ is the $n$-cycle. From this, we can easily calculate the R\'enyi entropies as 
\begin{equation}
    \log \Tr \rho_A^n = -\frac{c}{6}\left(n - \frac{1}{n}\right) \log \frac{l_A}{\epsilon} \implies \mathcal{D}(\rho_A) = \log \Tr \rho_A^3 - 2 \log \Tr \rho_A^2 = \frac{c}{18} \log \frac{l_A}{\epsilon},
\end{equation}
which is the first line of Eq. (\ref{eq:n3-cft2}). 

Coming to the reflected term, our $\rho_{AA^\star}$ is built from $\ket{\rho_{AB}}$, which is the $m=2$ member of the family $\ket{\rho_{AB}^{m/2}}$ of Ref.~\cite{Dutta:2019gen}. Its $n$-th moment requires $r = mn = 2n$ replicas with
\begin{equation}\label{eq:refl-entr-perms}
\begin{split}
    \sigma_A &= (1\,2)(3\,4)\cdots(2n{-}1\;2n),\\
    \sigma_B &= (1\;2n)(2\,3)\cdots(2n{-}2\;\,2n{-}1),
\end{split}
\end{equation}
and $\sigma_C = e$, as stated in the main text for $n = 3$. The twist at $-l_A$ is $\mathcal{T}_{\sigma_A}$, with $n$ two-cycles and dimension $nh(2)$, where we define 
\begin{equation}
    h(k) = \frac{c}{24}\left(k - \frac{1}{k}\right). 
\end{equation}
The twist at $l_B$ is $\mathcal{T}_{\sigma_B^{-1}}$, with the same dimension. The twist at $0$ is $\mathcal{T}_{\sigma_A^{-1}\sigma_B}$. A short computation shows that $\sigma_A^{-1}\sigma_B$ consists of two $n$-cycles, one on the odd and one on the even replicas (for $n=3$, $\sigma_A^{-1}\sigma_B = (1\,5\,3)(2\,4\,6)$), so its dimension is $2h(n)$. The normalized moment is
\begin{equation}
    \Tr\rho_{AA^\star}^n = \frac{\langle\mathcal{T}_{\sigma_A}(-l_A)\,\mathcal{T}_{\sigma_A^{-1}\sigma_B}(0)\,\mathcal{T}_{\sigma_B^{-1}}(l_B)\rangle}{\langle\mathcal{T}_{\sigma_2}(-l_A)\,\mathcal{T}_{\sigma_2^{-1}}(l_B)\rangle^n},
\end{equation}
where the denominator is $(\Tr\rho_{AB}^2)^n$, with $\sigma_2 = (1\,2)$ acting on two replicas. Using Eq.~\eqref{eq:S-3pt} with $h_1 = h_3 = n\,h(2)$ and $h_2 = 2h(n)$, the numerator scales as $(l_Al_B)^{-4h(n)}(l_A+l_B)^{-4(n h(2) - h(n))}$, up to powers of $\epsilon$. The denominator scales as $(l_A+l_B)^{-4n h(2)}$. The dimension $n h(2)$ of the outer twists cancels, and
\begin{equation}
    \Tr\rho_{AA^\star}^n = C_{n}\left(\frac{l_A\,l_B}{(l_A+l_B)\,\epsilon}\right)^{-\frac{c}{6}\left(n - \frac1n\right)}.
\end{equation}
The reflected spectrum is that of a single interval of effective length $l_Al_B/(l_A+l_B)$. The value of $m$ enters only through the outer twists and the OPE coefficient $C_{n}$, so using $\ket{\rho_{AB}}$ instead of $\ket{\sqrt{\rho_{AB}}}$ changes only non-universal constants. At $n\to1$ this reproduces the reflected entropy $S_R = \frac{c}{3}\log\frac{l_Al_B}{(l_A+l_B)\epsilon} + \text{const}$~\cite{Dutta:2019gen}. Substituting these moments into 
\begin{equation}
    \mathcal{D}^R(\rho_{AB}) = \log \Tr \rho_{AA^{\star}}^3 - 2 \log \Tr \rho_{AA^{\star}}^2 = \frac{c}{18} \log \frac{l_A l_B}{(l_A + l_B)\epsilon}
\end{equation}
which is the second line in Eq. (\ref{eq:n3-cft2}) from the main text. 

Finally, coming to the negativity term, the permutations are $\sigma_A = \tau_{2n}$, $\sigma_B = \tau_{2n}^{-1}$, and $\sigma_C = e$. The twist operator at $0$ carries the permutation $\tau_{2n}^{-2}$ which has two cycles of length $n$. The scaling dimensions of the three twist operators are therefore, 
\begin{equation}
    h_1 = h_3 = h(2n), \quad h_2 = 2 h(n).  
\end{equation}
From the three-point function of a CFT, we get 
\begin{equation}
    N_{2n} = \Tr \left(\rho_{AB}^{\Gamma}\right)^{2n} \propto l_A^{-4 h(n)} \, l_B^{-4 h(n)} \, \left(l_A + l_B\right)^{-4 h(2n) + 4 h(n)} = \left(l_A l_B\right)^{-\frac{c}{6}\left(n-\frac{1}{n}\right)} \left(l_A + l_B\right)^{-\frac{c}{6}\left(n + \frac{1}{2n}\right)}.
\end{equation}
Therefore, we have 
\begin{equation}
\begin{split}
    \log N_6 & = -\frac{4c}{9} \log \frac{l_A}{\epsilon} -\frac{4c}{9} \log \frac{l_B}{\epsilon}  - \frac{19c}{36} \log \frac{l_A + l_B}{\epsilon}, \\ 
    \log N_4 & = - \frac{c}{4} \log \frac{l_A}{\epsilon} - \frac{c}{4}\log \frac{l_B}{\epsilon} - \frac{3c}{8} \log \frac{l_A + l_B}{\epsilon}, \\ 
    \log N_2 & = -\frac{c}{4} \log \frac{l_A + l_B}{\epsilon}. 
\end{split}
\end{equation}
Putting these together, we get the contribution from the negativity term to be
\begin{equation}
    \mathcal{D}^N(\rho_{AB}) = \frac{c}{18} \log \frac{l_A}{\epsilon} + \frac{c}{18} \log \frac{l_B}{\epsilon} - \frac{c}{36} \log \frac{l_A + l_B}{\epsilon} = \mathcal{D}^R(\rho_{AB}) + \frac{c}{36} \log \frac{l_A + l_B}{\epsilon}, 
\end{equation}
which is the third line of Eq. (\ref{eq:n3-cft2}) from the main text. 

\subsection{Single random tensor}
Let us now calculate $\mathcal{N}_3$ for a single random tensor with equal but large local bond dimensions to derive Eq. (\ref{eq:n3-random}). We take the state $\ket{\psi}_{ABC}$ to be Haar random with $d_A = d_B = d_C = d$, and ask how each term of $\mathcal{N}_3$ scales as $d \to \infty$. The Haar average of the $r^{\rm th}$ moment of $\ketbra{\psi}$ is a sum over permutations \cite{Hayden:2016cfa}: 
\begin{equation}
    \mathbb{E}\left[\ketbra{\psi}^{\otimes r}\right] = \frac{1}{[d^3]_r}\sum_{\tau\in S_r}\Pi(\tau),
\end{equation}
where $[D]_r = D(D+1)\cdots(D+r-1) \approx D^r(1 + O(r^2/D))$ is the rising factorial. Inserting this into the definition of a multi-invariant, and using $\Tr[\Pi(\sigma)^\dagger\Pi(\tau)] = d^{\chi(\sigma^{-1}\tau)}$ on each party, where $\chi$ counts cycles, gives 
\begin{equation}\label{eq:S-haar}
    \mathbb{E}\,\mathcal{Z}(\sigma_A,\sigma_B,\sigma_C) = \sum_{\tau\in S_r} d^{-f(\tau)}\big(1 + O(d^{-3})\big), \quad \text{where} \quad 
    f(\tau) = \sum_{X\in\{A,B,C\}} |\sigma_X^{-1}\tau|.
\end{equation}
Here $|g| = r - \chi(g)$ is the minimal number of transpositions needed to write $g$, so $|g^{-1}h|$ is the Cayley distance $d(g,h)$ of the main text. At large $d$, multi-invariants self-average, and $\mathbb{E}\log\mathcal{Z}\approx\log\mathbb{E}\mathcal{Z}$. The triangle inequality $|\sigma_X^{-1}\sigma_Y|\le|\sigma_X^{-1}\tau| + |\tau^{-1}\sigma_Y|$, summed over the three pairs, gives
\begin{equation}\label{eq:S-bound}
    f(\tau)\ \ge\ \frac12\sum_{X<Y}|\sigma_X^{-1}\sigma_Y|.
\end{equation}
Equality holds if and only if $\tau$ lies on a Cayley-graph geodesic between every pair of boundary permutations. We call such a permutation $\tau$ a \emph{median}. 

For $\Tr\rho_A^n$ we have $(\sigma_A,\sigma_B,\sigma_C) = (\gamma_n,e,e)$ and $f(\tau) = |\gamma_n^{-1}\tau| + 2|\tau|$. The minimum $f = n-1$ is attained only at $\tau = e$. The next contributions, with $f = n$, come from the $\binom{n}{2}$ transpositions, all of which lie on the geodesic from $e$ to $\gamma_n$. Hence
\begin{equation}
    \Tr\rho_A^n = d^{1-n}\Big[1 + \frac{n(n-1)}{2d} + O(d^{-2})\Big].
\end{equation}
From these moments, we get 
\begin{equation}
    \mathcal{D}(\rho_A) = \frac{3}{d} - \frac{2}{d} + O(d^{-2}) = \frac{1}{d} + O(d^{-2}),
\end{equation}
and similarly for $B$ and $C$.

Let $Z_n = \mathcal{Z}(\sigma_A,\sigma_B,e)$ with $\sigma_A$ and $\sigma_B$ as in Eq. (\ref{eq:refl-entr-perms}), so that $\Tr\rho_{AA^\star}^n = Z_n/Z_1^n$. The pairwise distances are $|\sigma_A| = |\sigma_B| = n$ and $|\sigma_A^{-1}\sigma_B| = 2n-2$, so the bound~\eqref{eq:S-bound} is $f\ge 2n-1$. For $n=1$, $\sigma_A = \sigma_B = (1\,2)$, and $\tau = (1\,2)$ saturates the bound, giving $Z_1 = \Tr\rho_{AB}^2 = d^{-1}(1 + O(d^{-1}))$. For $n\ge2$, a median would need $|\tau| = 1$ and $|\sigma_A^{-1}\tau| = |\sigma_B^{-1}\tau| = n-1$, i.e.\ a transposition that appears in both $\sigma_A$ and $\sigma_B$. There is no such transposition, so the bound cannot be saturated and $f\ge 2n$. The value $f = 2n$ is attained at $\tau = e$. For $n = 2$ it is also attained at $\tau = g_A$ and $\tau = g_B$, but for $n\ge 3$ these have $f = 3n-2 > 2n$. Enumerating $S_4$ and $S_6$ confirms that there are no other minimizers. Hence $Z_2 \approx 3d^{-4}$ and $Z_3\approx d^{-6}$, so
\begin{equation}
    \Tr\rho_{AA^\star}^2 \approx \frac{3}{d^2},\quad \Tr\rho_{AA^\star}^3 \approx \frac{1}{d^3},\quad \mathcal{D}^R(\rho_{AB}) = \log\frac{d}{9} + O(d^{-1}), 
\end{equation}
which is the second term in Eq. (\ref{eq:n3-random}). 

For $2n$-moments of the partial transpose, we have $(\sigma_A,\sigma_B,\sigma_C) = (\tau_{2n},\tau_{2n}^{-1},e)$ with pairwise distances $2n-1$, $2n-1$, and $ 2n-2$ respectively. The bound~\eqref{eq:S-bound} is $f \geq 3n-2$, and it is saturated exactly by the non-crossing pairings, i.e., products of $n$ disjoint transpositions that are non-crossing with respect to the cyclic order of $\tau_{2n}$~\cite{Dong:2021clv}. There are $C_{n}$ such pairings, where $C_p$ is the $p$-th Catalan number. Hence
\begin{equation}
    N_{2n} \approx C_{n}\, d^{\,2 - 3n},
\end{equation}
the moments of a semicircle distribution~\cite{Shapourian:2020mkc}. The powers of $d$ drop out in the negativity term of $\mathcal{N}_3$, leaving
\begin{equation}
    \mathcal{D}^N(\rho_{AB}) = \log\frac{C_3C_1}{C_2^2} = \log\frac{5\cdot1}{2^2} = \log\frac54,
\end{equation}
which completes the derivation of Eq.~\eqref{eq:n3-random}. This is why the negativity term in Eq. \eqref{eq:n3-random} is $\mathcal{O}(1)$ while the reflected term is large. 

\subsection{Random tensor networks on the hyperbolic disk}
We consider a random tensor network on a hyperbolic disk, truncated a small distance $\epsilon$ away from the conformal boundary to regulate hyperbolic distances, and with bond dimensions $D \to \infty$. Averaging each tensor replaces it with a sum over permutations. This maps it to a spin-model where the spins take values from the permutation group. Schematically, we get something like 
\begin{equation}\label{eq:S49}
    \mathbb{E} \, \mathcal{Z} = \sum_{\{\tau_x\}} \exp\left[-\log D  \sum_{\langle xy\rangle} |\tau_x^{-1} \tau_y|\right].
\end{equation}
The multi-invariant to be calculated decides the boundary conditions, and in the large bond-dimension limit, the problem reduces to finding the saddle point solution. In the continuum limit, a domain wall of length $l$ separating the domains $g$ and $h$ costs $\kappa |g^{-1}h| l$, and the saddle point contributes 
\begin{equation}
   \log  \mathbb{E} \mathcal{Z} = - \min E + \log (\# \rm{minima}), \quad \text{where} \quad E = \kappa \sum_{\rm{walls}} l |g^{-1}h|
\end{equation}
It is economical to let the walls be geodesics on the hyperbolic disk. We call $T = |g^{-1} h|$ the tension of a domain wall. 

We work in the upper-half-plane model of the hyperbolic plane, with metric
\begin{equation}
    ds^2 = \frac{du^2 + dw^2}{w^2},  
\end{equation}
and the disk is truncated at $w = \epsilon$. The regularized geodesic length from a bulk point $y = (u, w)$ to a boundary point $x$ is 
\begin{equation}
    \sigma_{\epsilon}(x, y) = \log \frac{(u-x)^2 + w^2}{w \epsilon},
\end{equation}
and the geodesic distance between two boundary points $x_1$ and $x_2$ is 
\begin{equation}
    L(x_1, x_2) = 2 \log \frac{|x_{12}|}{\epsilon},
\end{equation}
where $|x_{12}| = |x_1 - x_2|$. On the disk, $|x_{12}|$ is replaced by the chord length $z_{12}$ between the two boundary points, as in the main text.

Let us now discuss how $\kappa = c/6$ is fixed by matching the entanglement entropy of a single interval on the boundary with the CFT result. The saddle that dominates in the calculation of $\Tr \rho_A^n$ is a single domain wall of tension $n-1$ connecting the two end points of the boundary region $A$. Therefore, we get 
\begin{equation}
    \Tr \rho_A^n \sim e^{- \kappa (n-1) L_A} \implies S(A) = \kappa L_A, 
\end{equation}
which when matched with the CFT result 
\begin{equation}
    S(A) = \frac{c}{3} \log \frac{l_A}{\epsilon} 
\end{equation}
gives $\kappa = c/6$, as $L_A = 2 \log l_A/\epsilon$. Moreover, from the above result, we immediately see that the marginal anti-flatness terms are all zero: 
\begin{equation}
    \mathcal{D}(\rho_A) = \mathcal{D}(\rho_B) = \mathcal{D}(\rho_C) = 0.
\end{equation}

\subsubsection{Three adjacent arcs.}
We now turn to the reflected term in $\mathcal{N}_3$. The dominant bulk saddle in Eq. \eqref{eq:S49} is a trivalent junction as shown in Fig. \ref{fig:rtn-saddles}(b). If $y = (u, w)$ is the location of the junction, then the free energy to be minimized is the length of each arm times the corresponding wall tension. We can use an $\rm SL(2, \mathbb{R})$ transformation to put the three endpoints on the $w = 0$ boundary at $x = \pm 1, 0$. There are two domain walls with tension $n$, and one with tension $2n-2$, and we place the tension $n$ walls at $\pm 1$, and the tension $2n-2$ wall at $0$. From symmetry, we expect the junction point lies on the imaginary axis, i.e., $u = 0$. Therefore, the free energy to be minimized becomes only a function of $w$: 
\begin{equation}
    \mathcal{F}(w) = 2n \log \frac{w^2 + 1}{w \epsilon} + (2n-2) \log \frac{w}{\epsilon}.
\end{equation}
The minimization gives
\begin{equation}
    \frac{4n w^2}{w^2 + 1} = 2 \implies w^{\star} = \frac{1}{\sqrt{2n - 1}}.
\end{equation}
Substituting this back into the free energy gives the minimized free energy: 
\begin{equation}
    \mathcal{F}(w^{\star}) = 2n \log 2n - (2n-1) \log (2n-1) - (4n - 2) \log \epsilon. 
\end{equation}
At $n = 1$, the junction sits on the geodesic between $\pm 1$ and $\mathcal{F}(w^\star) = 2\log(2/\epsilon)$, which is the saddle for the normalization $\Tr\rho_{AB}^2$. Dividing by $(\Tr\rho_{AB}^2)^n$ removes the $2n \log 2$ in $2n\log 2n$ and leaves a UV divergence $(2n-2)\log\epsilon$, which is proportional to $n-1$ and therefore cancels in the anti-flatness term. Therefore, we have
\begin{equation}
    \log \Tr \rho_{AA^{\star}}^n = -\kappa j(n) + \cdots, 
\end{equation}
where $\cdots$ vanish in $\mathcal{D}^R$, and $j(n) = 2n \log(n) - (2n-1)\log(2n-1)$ is the function defined in Eq. (\ref{eq:j(n)}) from the main text. Therefore, we get 
\begin{equation}
    \mathcal{D}^R = \kappa(2j(2) - j(3)) = \frac{c}{6} \log \frac{2^8 5^5}{3^{12}}. 
\end{equation}

The negativity term in $\mathcal{N}_3$ is dominated by a saddle of the form in Fig. \ref{fig:rtn-saddles}(c). A median $\tau$ fills the triangle bounded by the three entanglement wedges, and its walls have tensions $|\sigma_A^{-1} \tau| = n - 1$ along $\gamma_A$, $|\sigma_B^{-1} \tau| = n - 1$ along $\gamma_B$, and $|\tau| = n$ along $\gamma_C$. This saddle therefore has an energy 
\begin{equation}
    \mathcal{F} = (2n-2) \left(\log \frac{l_A}{\epsilon} + \log \frac{l_B}{\epsilon}\right) + 2n \log \frac{l_C}{\epsilon},
\end{equation}
where $l_A$, $l_B$, and $l_C$ are the chord lengths subtended by $A$, $B$, and $C$. The junction saddle has exactly the same $\epsilon$ dependence, but it costs an additional finite term coming from the analog of $j(n)$ for the reflected term. Therefore, the $\tau$ saddle dominates. Since the energy is linear in $n$, only the degeneracy of the $\tau$ saddle matters, and we are left with 
\begin{equation}
    \mathcal{D}^N(\rho_{AB}) = \log \frac{5}{4}
\end{equation}
as for the single tensor. This completes the derivation of Eq. (\ref{eq:rtn1}). We now turn to the more interesting case of non-adjacent arcs considered in the main text. 

\subsubsection{Two disjoint arcs.} 

Let $A = [z_1, z_2]$, $B = [z_3, z_4]$, with $C_1 = [z_2, z_3]$, and $C_2 = [z_4, z_1]$, and let
\begin{equation}
    L_{ij} = 2 \log \frac{z_{ij}}{\epsilon}. 
\end{equation}
The cross ratio of these points is $x = z_{12}z_{34}/z_{13} z_{24}$. Since the four points lie on a circle, we also have $1- x = z_{14}z_{23}/z_{13}z_{24}$. The entanglement wedge of $AB$ is connected when $L_{12}+L_{34} > L_{23} + L_{14}$, i.e., $x > 1- x$, or $x > 1/2$ \cite{Headrick:2010zt}. In the connected phase, we have 
\begin{equation}
    I(A:B) = \kappa \left(L_{12} + L_{34} - L_{23} - L_{14}\right) = \frac{c}{3} \log \frac{x}{1-x}. 
\end{equation}

We work in the upper-half plane as before and set $(z_1, z_2, z_3, z_4) = (-b, -a, a, b)$ on the real line. The cross-ratio is then $x = (b-a)^2/(b+a)^2$. Let's first work out the negativity term. The boundary conditions are the $2n$-cycle $\tau_{2n}$ on $A$, $\tau_{2n}^{-1}$ on $B$, and the identity on $C_1$ and $C_2$. The free energy in the disconnected and connected phases of Figs. \ref{fig:rtn-two-arcs}(a) and \ref{fig:rtn-two-arcs}(c) is
\begin{equation}
    \mathcal{F}_{(a)} = (2n-1) (L_{12} + L_{34}), \quad \mathcal{F}_{(c)} = (n-1)(L_{12}+L_{34}) + n (L_{23}+L_{14}), 
\end{equation}
when a median $\tau$ fills the interior region. This saddle has a degeneracy of $C_{n}$. The difference between the two energies is proportional to $L_{12}+L_{34} - L_{23} - L_{14}$ which changes sign precisely for $x = 1/2$ for every value of $n$. For $x < 1/2$, there is no degeneracy while for $x > 1/2$ there is a Catalan number of degenerate saddles. Therefore, we get 
\begin{equation}
    \mathcal{D}^{N} = \log \frac{5}{4} \, \Theta(x -1/2). 
\end{equation}
which is Eq. (\ref{eq:DN-rtn}) from the main text. 

Let's now consider the reflected term. As for the negativity term, there are two competing saddles: the one in Fig. \ref{fig:rtn-two-arcs}(a) and the $t$-channel saddle in Fig. \ref{fig:rtn-two-arcs}(b). The energy of the disconnected saddle is simply 
\begin{equation}
    \mathcal{F}_{(a)} = n(L_{12} + L_{34}), 
\end{equation}
as the domain wall tension is $n$. Coming to the $t$-channel network, we must once again minimize but now over the two junctions. From symmetry, these two junctions lie on the imaginary axis, say $y_1 =  (0, w_1)$ and $y_2 = (0, w_2)$. The middle wall then has a length $\log(w_2/w_1)$, and the free energy to be minimized is 
\begin{equation}
    \mathcal{F}(w_1, w_2) = 2n \log \frac{a^2 + w_1^2}{w_1 \epsilon} + 2n \log \frac{b^2 + w_2^2}{w_2 \epsilon} + (2n-2) \log \frac{w_2}{w_1}. 
\end{equation}
The two minimizations decouple, and we get 
\begin{equation}
    w_1^2 = (2n-1) a^2, \quad w_2^2 = b^2/(2n-1).
\end{equation}
This network exists only when $w_1 < w_2$, i.e., $b/a > 2n-1$. This condition holds at the transitions found below, where $b/a \approx 13$ at $x_2$ and $b/a \approx 20$ at $x_3$.
Substituting these back into the energy, we find 
\begin{equation}
    \mathcal{F}_{\min}(w_1, w_2) = n L_{23} + n L_{14} + 2j(n) + (2n-2) \log \frac{b}{a}.
\end{equation}

The disconnected saddle switches to the connected saddle in $\Tr \rho_{AA^{\star}}^n$ when 
\begin{equation}
    2j(n) + (2n-2) \log \frac{b}{a} = n\left(L_{12} + L_{34} - L_{23} - L_{14}\right) = 2n \log \frac{x}{1-x}.
\end{equation}
From $x = (b-a)^2/(b+a)^2$, we have $b/a = (1+\sqrt{x})/(1-\sqrt{x})$, and the transition point can be determined numerically. We find $x_2 \approx 0.735$, and $x_3 \approx 0.821$. Putting all of this together, we find the following four regimes: 
\begin{itemize}
    \item For $x < 1/2$, every moment is dominated by the disconnected saddle, and we have $\mathcal{D}^{R} = 0$. 

    \item For $1/2 < x < x_2$, the normalization is connected, but the numerators are disconnected for both $n = 2$ and $n =3$. Therefore, we find $\mathcal{D}^R = I(A:B)$. 

    \item For $x_2 < x < x_3$, the numerator is in the $t$-channel for $n = 2$ but disconnected for $n = 3$. Therefore, we have 
    \begin{equation}
        \mathcal{D}^R = \frac{2c}{3}\left(j(2) + \log \frac{1+\sqrt{x}}{1-\sqrt{x}}\right) - c \log \frac{x}{1-x}
    \end{equation}
    which decreases monotonically from $I(A{:}B)$ at $x_2$ to the plateau value below at $x_3$, since its derivative is $\frac{c}{x(1-x)}\left(\frac{2\sqrt{x}}{3} - 1\right) < 0$.

    \item For $x > x_3$, the numerators are in the $t$-channel for both $n = 2$ and $n = 3$, and we find
    \begin{equation}
        \mathcal{D}^R = 2 \kappa \left(2j(2) - j(3) \right) = \frac{c}{3} \log \frac{2^8 5^5}{3^{12}}, 
    \end{equation}
    exactly twice the value found for the adjacent arcs. 
\end{itemize}
This proves Eq. (\ref{eq:DR-two-arcs}) and gives the curve in Fig. \ref{fig:rtn-DR}.

\end{document}